\documentclass[a4paper,12pt]{article}
\usepackage[T1]{fontenc}
\usepackage[cp1250]{inputenc}
\usepackage[polish,english]{babel}
\usepackage[cp1250]{inputenc}
\usepackage{graphicx}

\title{Comments on Evolution of Cooperation: 
Assortative Encounters with Selfknowledge}
\author{Paweł Sobkowicz\footnote{e-mail address: {\tt pawelsob@poczta.onet.pl}}}
\date{9th October 2003}

\begin{document}

\maketitle

\section{Abstract}

The theoretical description of the evolution of cooperation presented by Bergstrom~\cite{berg1}  based on assortative matching with partner choice allows to model the population dynamics in a game of Nonrepetitive Prisoners Dilemma. \\
In this paper we present a short analysis of asymmetric effects brought into the game by self knowledge of the participants, that is the knowledge of one's own strategy.  Within the same conceptual framework as introduced by Bergstrom inclusion of selfknowledge leads to different behaviour of the assortativity index and changes in payoffs  for different strategies.  

\section{Basic Assumptions and Notation}

Since the classic works by Maynard Smith \cite{maynard73,maynard82} the evolutionary game theory has been used in numerous applications, ranging from biology, through sociology to applied economics. One of the branches of the discipline is the matching theory, which deals with the situations where two separate sets of players have to with each other. Matching theory can describe fields as sexual pairing, job market or even routing in computer networks. The literature dealing with matching theory is quite rich, the introduction can be found in the work of Roth and Sotomayor~\cite{roth82}. 

One of the interesting aspects of matching is the influence of intergroup traits on the matching process. Suppose that there are some subsets  within each of the two sets of players that are matched. the question whether it makes sense to include these subsets in analysis of the matching, or, in other words, whether there is any evolutionary advantage in matching like with like (or vice versa) has been given the name of assortativity matching. This paper is a direct extension of the analysis of assortative matching presented by Bergstrom~\cite{berg1}. We would repeat a lot of the definitions and notations used in that paper to allow to follow the reasoning and easily observe the differences of the two approaches.

Following Bergstrom we assume that our population is divided between two types of players, \emph{cooperators} and \emph{defectors}. The ratio of cooperators within the population is $x$, correspondingly for defectors it is $1-x$.  

The members of the population form pairs (match) with other members. Assortativity of matching results from difference between of frequencies of matching between one's own type and with different type. 
The relative frequency of matchings between various types of players depends on individuals' types and proportions of their types within the population.

Let $p(x)$ be the conditional probability that one encounters a cooperator, \emph{given that one is a cooperator}.\\
Let $q(x)$ be the conditional probability that one encounters a cooperator, \emph{given that one is a defector}.

Then, as argued in \cite{berg1}, the fraction of all encounters between two individuals in which a cooperator meets a defector is $x(1-p(x))$. Similarly, the fraction of encounters in which a defector meets a cooperator is $(1-x)q(x)$. Since these are just two different ways of counting the same matchings we have the initial equation:
\begin{equation}
\label{parity}
x(1-p(x)) = (1-x)q(x)
\end{equation}

Bergstrom defined the \emph{index of assortativity}, $a(x)$ as a difference between probability of homogenous and heterogenous matches.
\begin{equation}
\label{ax}
a(x) = p(x)- q(x)
\end{equation}

By simple rearangements it is possible to express $p(x)$ and $q(x)$  through $x$ and $a(x)$:
\begin{equation}
q(x) = x [ 1- (p(x)-q(x)] = x (1-a(x))
\end{equation}

\begin{equation}
p(x) = a(x) + x(1-a(x))
\end{equation}

\subsection{Basic games scenarios}

Let's define the general payoff matrix for two participants, $X$ and $Y$. Each of them has two possible choices of behaviour, Cooperating ($C$) and Defecting ($D$). The payoffs are defined as:
\begin{center}
\begin{table}[h]
\begin{tabular}{|c|c|c|} \hline
      & $Y=C$   &  $Y=D$ \\ \hline
$X=C$ &  $R,R$  &  $S, T$ \\ \hline
$X=D$ &  $T,S$  &  $P, P$ \\ \hline
\end{tabular}
\end{table}
\end{center}

For the games considered here we assume that $T \ge R \ge P \ge S$. 

In this paper we would be assuming that the individual players use simple strategies (i.e. not resorting to adaptive or time-dependant strategies, such as Repeated Prisoners Dilemma (RPD) games). Moreover, we would assume that a player would accept a match when the expected payoff is greater than or equal zero (assumption of rational behaviour). This means that if the analysis, based on the information available to the player would show that he is certain to lose, the player would not match. This point would become important when we would start discussing the assortative matching with partner choice.

Bergstrom has considered two types of games: prisoners Dilemma and the Game of Shared Output. The analysis of assortativity applies to both of them, but for simplicity we would concentrate on the game of  Prisoners Dilemma. Within this game the payoff for player X is given by:
\begin{description}
\item[Both coooperate:] $R = b-c $
\item[$X$ cooperates, $Y$ defects:] $S = -c $  ($ S<0$!)
\item[$X$ defects, $Y$ cooperates:] $T = b$
\item[Both defect:] $P=0$
\end{description}
where $b$ is the benefit conferred and $c$ is the cost of action, with $b>c$.

\section{Matching rules}

As Bergstrom has noted, when partners have choice about their partners and matching results from mutual consent interesting possibilities arise. This allows us to predict the index of assortativity. In the game of Prisoners Dilemma everyone would prefer to be matched with a cooperator rather than with a defector. In the simplest case, if the players behaviour could be predicted by observation with a 100\% accuracy then \emph{the cooperators would match with cooperators only (with payoff of $R$)and defectors only with defectors  (with payoff of $P=0$)}. In this case, regardless of $x$, $p(x) = 1$, $q(x)=0$ and, consequently $a(x) = 1$. With $R>P$ the only stable population would be the one of all cooperators.

Bergstrom proposed a more general model, in which the accuracy of determination of partner type of behaviour is less than perfect. The proposed  model is based on \emph{labelling} the players with an imperfect indicator of their type (such a s reputation based on partial information, behavioral cues or a psychological test \cite{berg1}). We introduce notation in which, for example, $D_C$ denotes a true defector labelled and perceived as cooperator (good cheater),
$C_C$ denotes true and recognizable cooperator, $C_D$ a cooperator unfortunately mislabelled as defector and $D_D$ a defector unable to 'hide' his character.

Within the population of cooperators we have a fraction $\alpha$ of correctly labelled ones ($C_C$) and fraction $(1-\alpha)$ of cooperators mislabelled as defectors $C_D$. Within the population of defectors we would have a fraction 
$\beta$ of correctly labelled $D_D$ and fraction $(1-\beta)$ of defectors mislabelled as cooperators ($D_C$). Bergstrom proposed that labelling should be `reasonably reliable', i.e. $\alpha, \beta > 1/2$, we would study the whole range of 
$0<\alpha, \beta < 1$.

Because the decisions of the players would be based on what they assume about the other players (the labels) rather than the real situation the outcome of the matching (payoff matrix) becomes less trivial.

Bergstrom assumed a crucial simplification in partner acceptance rules, in which \emph{`apparent cooperators will all be matched with apparent cooperators and apparent defectors would be matched with apparent defectors'}. This allowed to calculate the index of assortativity for any $x$.

The assumption above does not, however, apply to the situation when the players are selfknowledgeable, that is when any player `knows' his `true nature'. In such situation true cooperators -- even those labelled as defectors --- would shirk from matching with perceived defectors. This would create quite complicated mutual consent rules and payoff matrix which are  presented in several tables in Appendix~1.
The cases where the mutual consent is marked by YES$^*$ are special in the sense of additional level of selfknowledge, possible to be added, namely the knowledge of other's reaction to one's own label. For example in the case $X = C_D$ and $Y=D_C$,  $X$, knowing he looks like a defector might wonder why an apparent cooperator would want to match with him. The only rational explanation being that $Y$ is in fact a true defector, $X$ would then rationalise \emph{not} to match with $Y$. Such in depth analysis of 'I know that you know that I know \ldots ' is beyond the scope of the current analysis.

\subsection{Calculation of relative frequencies and assortative index}

In this section we will compare the calculated values of $p(x)$, $q(x)$ and $a(x)$ for both the Bergstrom model and Selfknowledge model.

Following {Bergstrom}  we have:

\begin{equation}
p_B(x) = \frac{\alpha^2 x}{\alpha x +(1-\beta)(1-x)} + \frac{(1-\alpha)^2 x}{(1-\alpha)x + \beta (1-x)}
\end{equation}

\begin{equation}
q_B(x) = \frac{(1-\beta) \alpha x}{\alpha x +(1-\beta)(1-x)} + \frac{\beta (1-\alpha) x}{(1-\alpha)x + \beta (1-x)}
\end{equation}

\begin{equation}
a_B(x) = (\alpha+\beta -1)\left(\frac{\alpha x}{\alpha x +(1-\beta)(1-x)} + \frac{(1-\alpha) x}{(1-\alpha)x + \beta (1-x)}\right)
\end{equation}

For the selfknowledge model the probabilities are:

\begin{equation}
p_S(x) = \frac{\alpha^2 x}{\alpha x +(1-\beta)(1-x)} 
\end{equation}

\begin{equation}
q_S(x) = (1-\beta) x
\end{equation}

\begin{equation}
a_S(x) = \frac{\alpha^2 x}{\alpha x +(1-\beta)(1-x)} - (1-\beta) x
\end{equation}

Figure 1 in Appendix 2 compares $a_B(x)$ and $a_S(x)$ for $\alpha = 3/4$ and $\beta = 3/5$. One can immediately see the difference in behaviour of $a(x)$ in the region of $x\approx 1$. To understand the meaning of the differences it is useful to consider the limiting cases for both models.

\begin{table}[!h]
\begin{tabular}{|p{6.8cm}|p{6.8cm}|} \hline
{\bf Bergstrom} & {\bf Selfknowledge} \\ \hline \hline
\multicolumn{2}{|c|}{$p(x)$ in the limit $x \rightarrow 0$} \\ \hline
$p_B(x) \approx x\left(\frac{\alpha^2}{1-\beta} + \frac{(1-\alpha)^2}{\beta}\right)$ & 
$p_S(x) \approx x\left(\frac{\alpha^2}{1-\beta} \right)$ \\ \hline
\multicolumn{2}{|c|}{$q(x)$ in the limit $x \rightarrow 0$} \\ \hline
$q_B(x) \approx x$ & $q_S(x) \approx x(1-\beta)$ \\ \hline
\multicolumn{2}{|c|}{$p(x)$ in the limit $x \rightarrow 1$} \\ \hline
$p_B(x) \approx 1 - (1-x) $ & $p_S(x) \approx \alpha - (1-x)(1-\beta) $ \\ \hline
\multicolumn{2}{|c|}{$q(x)$ in the limit $x \rightarrow 1$} \\ \hline
$q_B(x) \approx 1 - (1-x)\left(\frac{\beta^2}{1-\alpha}+\frac{(1-\beta)^2}{\alpha}\right) $ 
& $q_S(x) \approx (1-\beta) - (1-x)(1-\beta) $ \\ \hline
\end{tabular}
\end{table}

The intuitive explanation of the limiting cases is quite instructive. Lets consider first the $x\rightarrow 0$. In Bergstrom's model $p_B(x)$ is proportional to $x$, with two terms corresponding to two situations: real cooperator laballed as cooperator matches real cooperator laballed as cooperator and real cooperator laballed as defector matches with real cooperator laballed as defector. Taking selfknowledge into account the second type of match is not allowed, as selfknowledgeable cooperator would not match with apparent defector, thus $p_S(x)$ has only one term.

For $q(x)$ in the limit of small $x$ , Bergstrom's model gives $q_B(x)\approx x $. This corresponds to the situation where every defector (either $D_C$ or $D_D$) would find a cooperator (respectively $C_C$ and $C_D$). In the selfknowledge model \emph{only} the defectors posing as cooperators would find willing cooperators, and thus $q_S(x) \approx x(1-\beta) $.

Despite quantitative differences both model preserve linaerity of $p(x)$ and $q(x)$ for small $x$. For $x$ approaching 1 their behaviour is however drastically different. Bergstrom's model predicts both functions to approach 1, so that their difference, $a(x)$ is again linear in $(1-x)$. The selfknowledge model gives a different prediction. As $x \rightarrow 1$ $p_S(x)$ approaches $\alpha$. This corresponds to an intuition that when there are almost no defectors, only the apparent cooperators would find willing partners. Those 'unfortunate' to be labelled as defectors, despite their true nature would remain unmatched. For $x \rightarrow 1$ we have $q_S(x) \rightarrow 1-\beta $. This may be explained as follows: from the small number of defectors \emph{all} cheaters labelled as cooperators would surely find a match of a cooperator. More importantly, the linear terms in expansion of both $  p_S(x)$ and $q_S(x)$ for $x$ approaching 1 are the same, namely 
$(1-x)*(1-\beta)$. They cancel out leaving $a_S(x) \approx \alpha +\beta -1 + {\cal O}(x^2) $.

\section{Comparison of payoffs}

Before we would summarize the results of the discussed models lets introduce the payoff functions for cooperators and defectors and their difference $\delta(x)$. This difference  would then determine the dynamics of the population. 

The payoff for a cooperator is given by:
\begin{eqnarray}
p(x) R + (1-p(x)) S & = & S + p(x)(R-S) \\ \nonumber
  & = & S + a(x)(R-S) + x(1-a(x))(R-S).
\end{eqnarray}
Similar reasoning gives the payoff for a defector:
\begin{equation}
q(x)T+(1-q(x))P = P + x(1-a(x))(T-P).
\end{equation}

The difference between payoffs is
\begin{equation}
\delta(x) = S-P +a(x)(R-S) + x(1-a(x))[(R+P) - (S+T)].
\end{equation}
For additive Prisoners Dilemma games $(R+P) - (S+T) = 0$ and
\begin{equation}
\delta(x) = S-P +a(x)(R-S) = a(x)b -c,
\end{equation}
where we have used the notation from Section 2. In the following we would study only such additive games.

In Bergstrom's model $a_B(x)$ for all values of $\alpha$ and $\beta$ has zero values for $x=0$ and $x=1$. For the range of values considered in \cite{berg1} i.e. $\alpha, \beta >1/2$, the index of assortativity $a_B(x)$ is a concave function. It is interesting to observe that this property is preserved for all values of $\alpha$ and $\beta$, with the exception of situation when $\alpha + \beta = 1$\footnote{For the case of $\alpha + \beta = 1 $ we have $a_B(x) \equiv 0 $ for all 
$x$}. This allows to extend the results of Bergstrom to quite interesting situation of invasion of 'very good cheaters' into cooperators population. In such situation $x \approx 1$,  $\beta < 1/2 $ and changing $\alpha$ does not introduce qualitative impact. As may be seen from Figure~2 even for $\beta < 1/2 $ the concave character of the $a_B(x)$ is preserved. This means that the concave character of $\delta_B(x) =a_B(x) b -c $ is also preserved. If $c> \max(a(x) b)$ then $\delta_B(x) <0$ and proportion of defectors increases, due to their higher payoff. There is only one stable point: all players are defectors. For $c < \max(a(x) b)$ as presented in \cite{berg1} $\delta_B(x) <0$ for certain $ x_1 <x <x_2$, and there are two stable equilibria: $x=0$ -- \emph{all defectors} and $x=x_2$ -- \emph{mixed population of defectors and cooperators} (see Figure 4).

In the selfknowledge model the situation is different. For certain values of $b, c$, such as $c> \max(a(x) b)$, as before, the proportion of defectors increases, due to their higher payoff. There is only one stable point: all players are defectors. 

When $\alpha +\beta -1 < 0$ we have $a_S(x) < 0 $ for all $x$. In such a case, regardless of the payoff coefficients $b$ and $c$ defectors benefit always surpasses cooperators benefit  and $\delta_S(x) <0$ for all $x$. The only equilibrium is at $x=0$ (see Figure 5).

When $\alpha +\beta -1 > 0$ the assortativity index increases monotonically with $x$, and for $ c <  \max(a(x) b)$ there is a region of $x>x_1$ where $\delta_S(x) > 0$. Two stable equilibria form depending on initial parameters, one at $x=0$ and one at $x=1$.

Interpretation of the above results based on common sense reasoning is as follows.

For $\alpha + \beta \ll  1$ we have relatively large number of `mislabelled' players, both cooperators and defectors. In such situation cooperators are at disadvantage: they would not want to match with anyone `looking like' a defector decreasing the number of matches with true cooperators, while still being prone to fall for a cheating defector. Defectors, on the other hand, see little to lose in matching with presumed defector and willingly match with presumed cooperators. Thus the negative value of $a(x)$ and following it the greater payoff for defectors.

For $\alpha +\beta \gg 1$ there are relatively few mislabelled characters. For large $x$, as most of the small number of defectors look as defectors, the pure strategy of cooperators  is actually beneficial. There are no matches such as $C_D$ (cooperator labelled as defector) willing to match with $D_D$ (correctly labelled defector) --- and such matches were allowed in Bergstrom model. Thus the difference in the predictions of the two models. The defectors matching with defectors are at payoff disadvantege to cooperators matching with cooperators and eventually vanish from population.

On the other hand, for small $x$ and $\alpha +\beta \gg 1$ the few cooperators present, surrounded by many correctly labelled defectors can match only within their group ($\alpha x$) or with the mislabelled defectors $(1-\beta)(1-x)$ --- the other situation resulting in severly disadvantageous outcome. Here the only stable population is that of defectors only.

A few words should be said about the validity of the two models for $\alpha +\beta \approx  1$. For both models this choice of parameters yields $a(x) \approx 0$ for all $x$. Moreover, the function $a(x)$ is very flat, and for $\alpha +\beta = 1$ this function is equal to 0 for all $x$. the validity of predictions for both models are questionable for such case and the dynamics of the population should be determined by factors not taken into account.

\subsection{Conclusions}

In this short analysis we have extended the theoretical model proposed by Bergstrom, which allowed not only to analyse the results of assortative matching (i.e. preferable matching of similar players among themselves) but also to calculate the assortativity index $a(x)$. The new model takes into account influence of the selfknowledge of the players on their choices. It has been shown that for certain values of the population composition (defined by numbers of cooperators and defectors and proportions of these populations `correctly' labelled) the stable points predicted by Bergstrom's model and the selfknowledge model differ qualitatively, the latter model leading only to `pure' populations of all defectors or all cooperators.

\newpage
\section*{Appendix 1: Matching tables for Selfknowledge model and for the Bergstrom model}

\subsection*{Tables for the Selfknowledge model}

\begin{table}[h]
\small{
\begin{tabular}{|c|c|c|c|c|c|p{5cm}|} \hline
\multicolumn{6}{|c}{Cooperator -- Cooperator}& {\it with selfknowledge}\\ \hline
$X$ & $Y$ & Fraction & Consent & Payoff $X$ & Payoff $Y$ & Reasoning \\ \hline
$C_C$  & $C_C$ & $\alpha^2 x^2 $ & YES   & $b - c +k $& $b - c +k $  & Full agreement of perceived and real behaviour \\ \hline
$C_C$  & $C_D$ & $\alpha  (1-\alpha)x^2 $ & NO   & --- & --- & $X$ would not agree to pair with $Y$, because $Y$ looks like $D$ \\ \hline
$C_D$  & $C_C$ & $(1-\alpha)\alpha x^2 $ & NO   & --- &---& $Y$ would not agree to pair with $X$, because $X$ looks like $D$  \\ \hline
$C_D$  & $C_D$ & $(1-\alpha)^2 x^2 $ & NO   & --- &--- & Neither would agree to pair with the other: 'I know I am a 
$C$, why would I pair with someone who looks like $D$' \\ \hline \hline
\end{tabular}
}
\end{table}

\begin{table}[h]
\small{
\begin{tabular}{|c|c|c|c|c|c|p{5cm}|} \hline
\multicolumn{6}{|c}{Defector -- Defector}& {\it with selfknowledge}\\ \hline
$X$ & $Y$ & Fraction & Consent & Payoff $X$ & Payoff $Y$ & Reasoning \\ \hline
$D_C$  & $D_C$ & $(1-\beta)^2 (1- x)^2  $ & YES   & 0 & 0 & Both would agree with hopes of cheating the other \\ \hline
$D_D$  & $D_D$ & $\beta^2  (1-x)^2 $ & YES   & 0 & 0 & Both would agree on principe of 'it would not hurt'\\ \hline
$D_D$  & $D_C$ & $\beta (1- \beta) (1-x)^2 $ & YES   & 0 & 0 & $X$ would agree with hope of cheating $Y$, $Y$ on principe of 'it would not hurt'\\ \hline
$D_C$  & $D_D$ & $ \beta (1- \beta) (1-x)^2 $ & YES   & 0 & 0& $Y$ would agree with hope of cheating $X$, $X$ on principe of 'it would not hurt'\\ \hline \hline
\end{tabular}
}
\end{table}

\begin{table}[h]
\small{
\begin{tabular}{|c|c|c|c|c|c|p{4cm}|} \hline
\multicolumn{6}{|c}{Cooperator -- Defector}& {\it with selfknowledge} \\ \hline
$X$ & $Y$ & Fraction & Consent & Payoff $X$ & Payoff $Y$ & Reasoning \\ \hline

$C_C$  & $D_C$ & $\alpha (1-\beta) x(1- x)  $ & YES   & $-c$ & $b$ & Both would agree, $X$ genuinely, $Y$ as cheater \\ \hline
$D_C$  & $C_C$ & $\alpha (1-\beta) x(1- x)  $ & YES   & $b$ & $-c$ & Both would agree, $Y$ genuinely, $X$ as cheater \\ \hline

$C_C$  & $D_D$ & $\alpha \beta x(1- x)  $     & NO   & --- & --- & $X$ would not agree \\ \hline
$D_D$  & $C_C$ & $\alpha \beta x(1- x)  $     & NO   & --- & --- & $Y$ would not agree \\ \hline

$C_D$  & $D_D$ & $(1-\alpha) \beta x(1- x)  $ & NO   & --- & --- & $X$ would not agree \\ \hline
$D_D$  & $C_D$ & $(1-\alpha) \beta x(1- x)  $ & NO   & --- & --- & $Y$ would not agree \\ \hline

$C_D$  & $D_C$ & $(1-\alpha) (1-\beta) x(1- x)  $ & YES$^*$   & $-c$ & $b$ & $X$ would agree perceiving $Y$ as $C$, $Y$ would agree in the spirit 'it would not hurt' \\ \hline
$D_C$  & $C_D$ & $(1-\alpha) (1-\beta) x(1- x)  $ & YES$^*$   & $b$ & $-c$ & $Y$ would agree perceiving $X$ as $C$, $X$ would agree in the spirit 'it would not hurt' \\ \hline
\end{tabular}
}
\end{table}

\subsection*{Tables for the Selfknowledge model}

\begin{table}[h]
\small{
\begin{tabular}{|c|c|c|c|c|c|p{5cm}|} \hline
\multicolumn{6}{|c}{Cooperator -- Cooperator}& {\it Bergstrom}\\ \hline
$X$ & $Y$ & Fraction & Consent & Payoff $X$ & Payoff $Y$ & Reasoning \\ \hline
$C_C$  & $C_C$ & $\alpha^2 x^2 $ & YES   & $b - c +k $& $b - c +k $  & apparent cooperator ONLY with apparent cooperator\\ \hline
$C_C$  & $C_D$ & $\alpha  (1-\alpha)x^2 $ & NO   & --- & --- & NO mixed labels\\ \hline
$C_D$  & $C_C$ & $(1-\alpha)\alpha x^2 $ & NO   & --- &---& NO mixed labels\\ \hline
$C_D$  & $C_D$ & $(1-\alpha)^2 x^2 $ & YES   & $b - c +k $   & $b - c +k $   & apparent defector ONLY with apparent defector\\ \hline \hline
\end{tabular}
}
\end{table}

\begin{table}[h]
\small{
\begin{tabular}{|c|c|c|c|c|c|p{5cm}|} \hline
\multicolumn{6}{|c}{Defector -- Defector}& {\it Bergstrom}\\ \hline
$X$ & $Y$ & Fraction & Consent & Payoff $X$ & Payoff $Y$ & Reasoning \\ \hline
$D_C$  & $D_C$ & $(1-\beta)^2 (1- x)^2  $ & YES   & 0 & 0 & apparent cooperator ONLY with apparent cooperator\\ \hline
$D_D$  & $D_D$ & $\beta^2  (1-x)^2 $ & YES   & 0 & 0 & apparent defector ONLY with apparent defector\\ \hline
$D_D$  & $D_C$ & $\beta (1- \beta) (1-x)^2 $ & NO   & --- & --- & NO mixed labels\\ \hline
$D_C$  & $D_D$ & $ \beta (1- \beta) (1-x)^2 $ & NO   & --- & --- & NO mixed labels\\ \hline \hline
\end{tabular}
}
\end{table}

\begin{table}[h]
\small{
\begin{tabular}{|c|c|c|c|c|c|p{4cm}|} \hline
\multicolumn{6}{|c}{Cooperator -- Defector}& {\it Bergstrom} \\ \hline
$X$ & $Y$ & Fraction & Consent & Payoff $X$ & Payoff $Y$ & Reasoning \\ \hline

$C_C$  & $D_C$ & $\alpha (1-\beta) x(1- x)  $ & YES   & $-c$ & $b$ & apparent cooperator ONLY with apparent cooperator\\ \hline
$D_C$  & $C_C$ & $\alpha (1-\beta) x(1- x)  $ & YES   & $b$ & $-c$ & apparent cooperator ONLY with apparent cooperator\\ \hline

$C_C$  & $D_D$ & $\alpha \beta x(1- x)  $ & NO   & --- & --- & NO mixed labels\\ \hline
$D_D$  & $C_C$ & $\alpha \beta x(1- x)  $ & NO   & --- & --- & NO mixed labels\\ \hline

$C_D$  & $D_D$ & $(1-\alpha) \beta x(1- x)  $ & YES   & $-c$ & $b$ & apparent defector ONLY with apparent defector\\ \hline
$D_D$  & $C_D$ & $(1-\alpha) \beta x(1- x)  $ & YES   & $b$ & $-c$ & apparent defector ONLY with apparent defector\\ \hline

$C_D$  & $D_C$ & $(1-\alpha) (1-\beta) x(1- x)  $ & NO   & --- & --- & NO mixed labels\\ \hline
$D_C$  & $C_D$ & $(1-\alpha) (1-\beta) x(1- x)  $ & NO   & --- & --- & NO mixed labels\\ \hline

\end{tabular}
}
\end{table}

\clearpage
\section*{Appendix 2: Figures}
\clearpage

\begin{figure}[ph]
\centering
\includegraphics{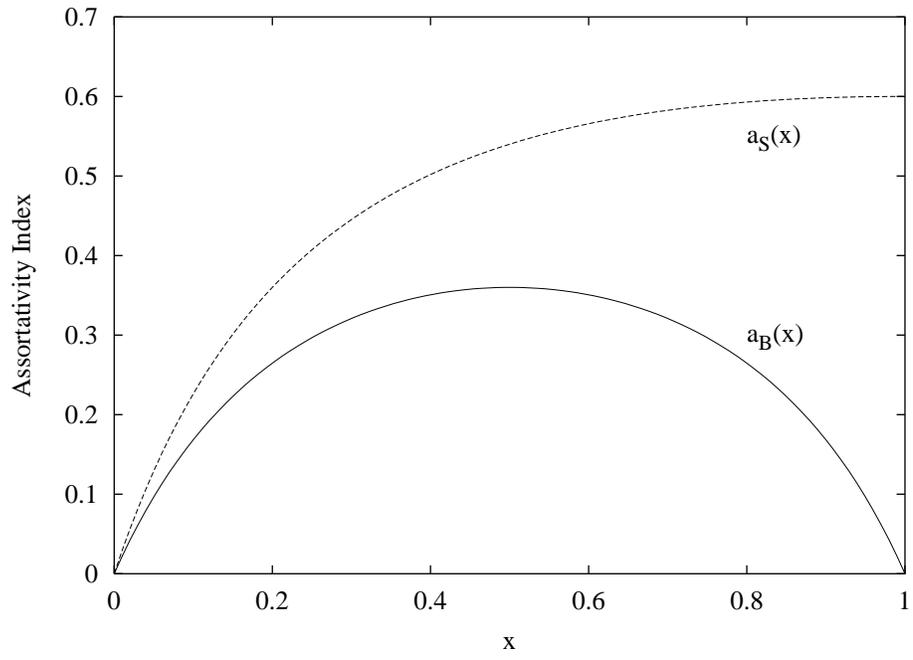}
\caption{Comparison of assortativity indexes for Bergrstrom model and for model with selfknowledge. Values used in calculation: $\alpha = 3/4$, $\beta = 3/5$. }
\end{figure}

\begin{figure}[ph]
\centering
\includegraphics{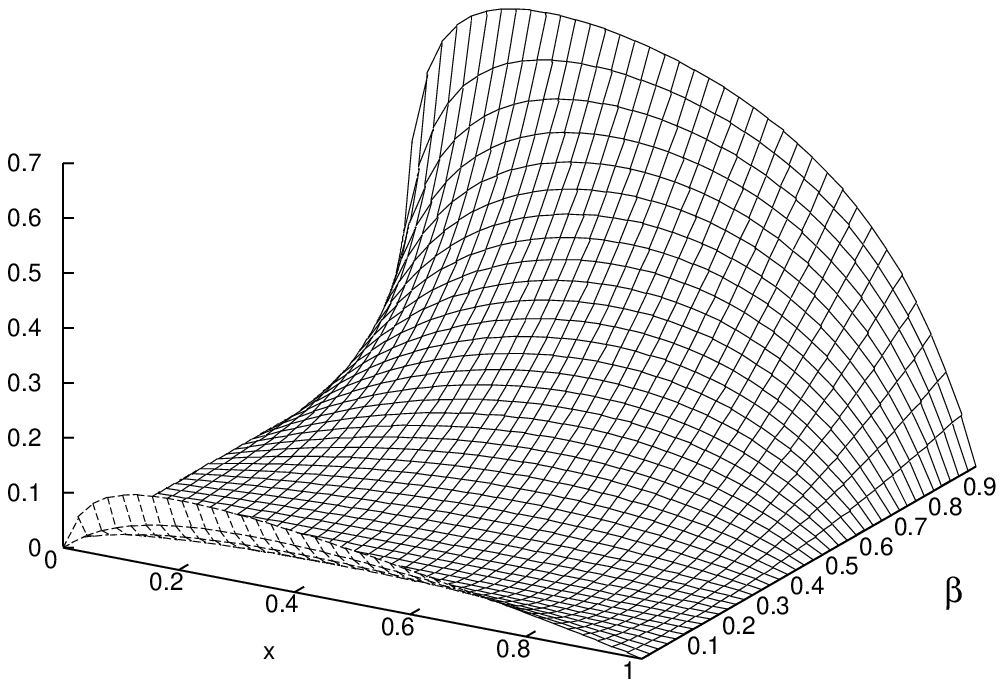}
\caption{Assortativity index $a_B$ for Bergrstom model as a function of $x$ and $\beta$. Values used in calculation: $\alpha = 0.8$; it is worth noting that for $\beta=0.2$ $a_B\equiv 0$. }
\end{figure}

\begin{figure}[ph]
\centering
\includegraphics{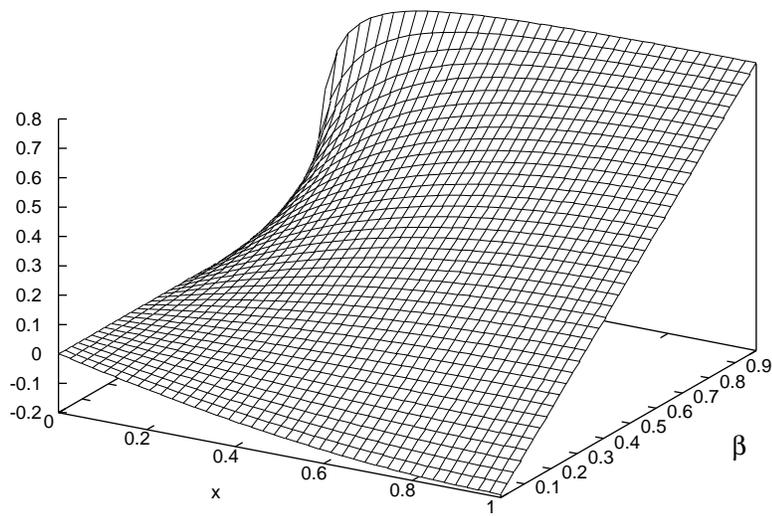}
\caption{Assortativity index for Selfknowledge model as a function of $x$ and $\beta$. Values used in calculation: $\alpha = 0.8$; it is worth noting that for $\beta=0.2$ $a_S\equiv 0$. }
\end{figure}

\begin{figure}[ph]
\centering
\includegraphics{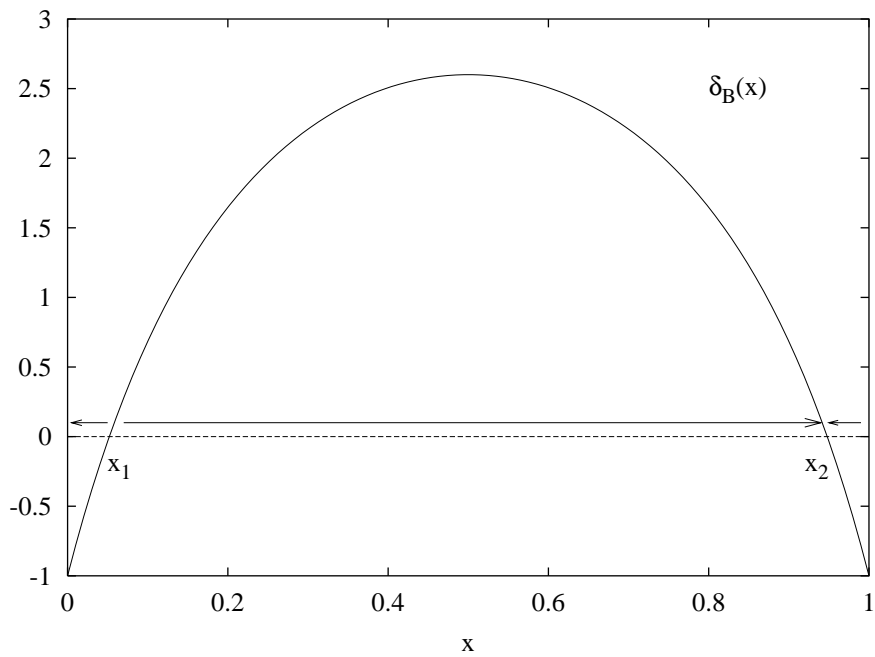}
\caption{Example of difference between payoffs $\delta(x)$ for Bergstrom model. Two stable equilibria at 
$x =0$ and $x= x_2$ form. Values used in calculation: $\alpha = 0.8$, $\beta=0.8$, $b=10$, $c=1$. Arrows indicate direction of evolution of populations. }
\end{figure}

\begin{figure}[ph]
\centering
\includegraphics{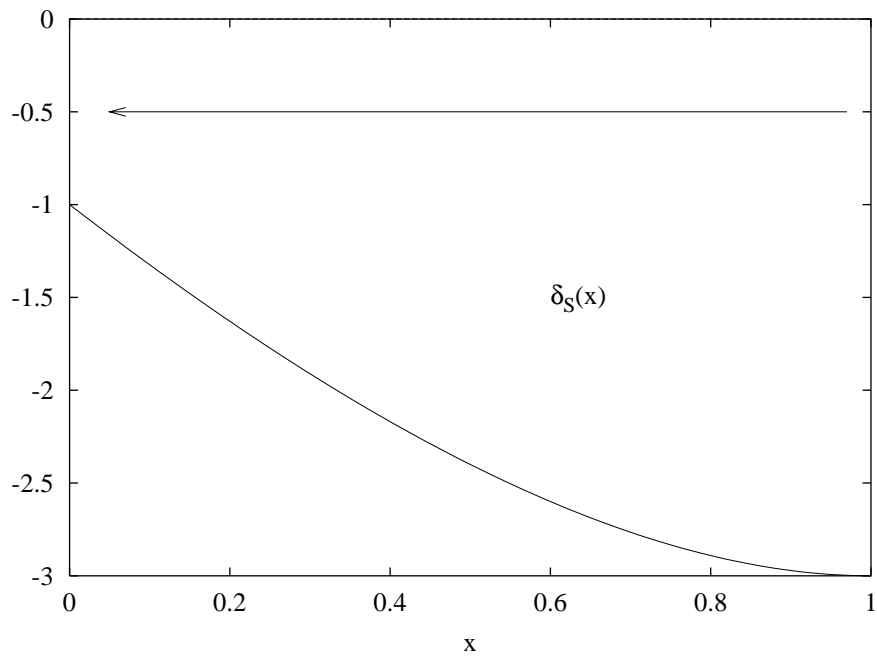}
\caption{Example of difference between payoffs $\delta(x)$ for Selfknowledge model with $\alpha + \beta < 1$ . Only one stable equilibrium at 
$x =0$ is present. Values used in calculation: $\alpha = 0.4$, $\beta=0.4$, $b=10$, $c=1$. Arrows indicate direction of evolution of populations. }
\end{figure}

\begin{figure}[ph]
\centering
\includegraphics{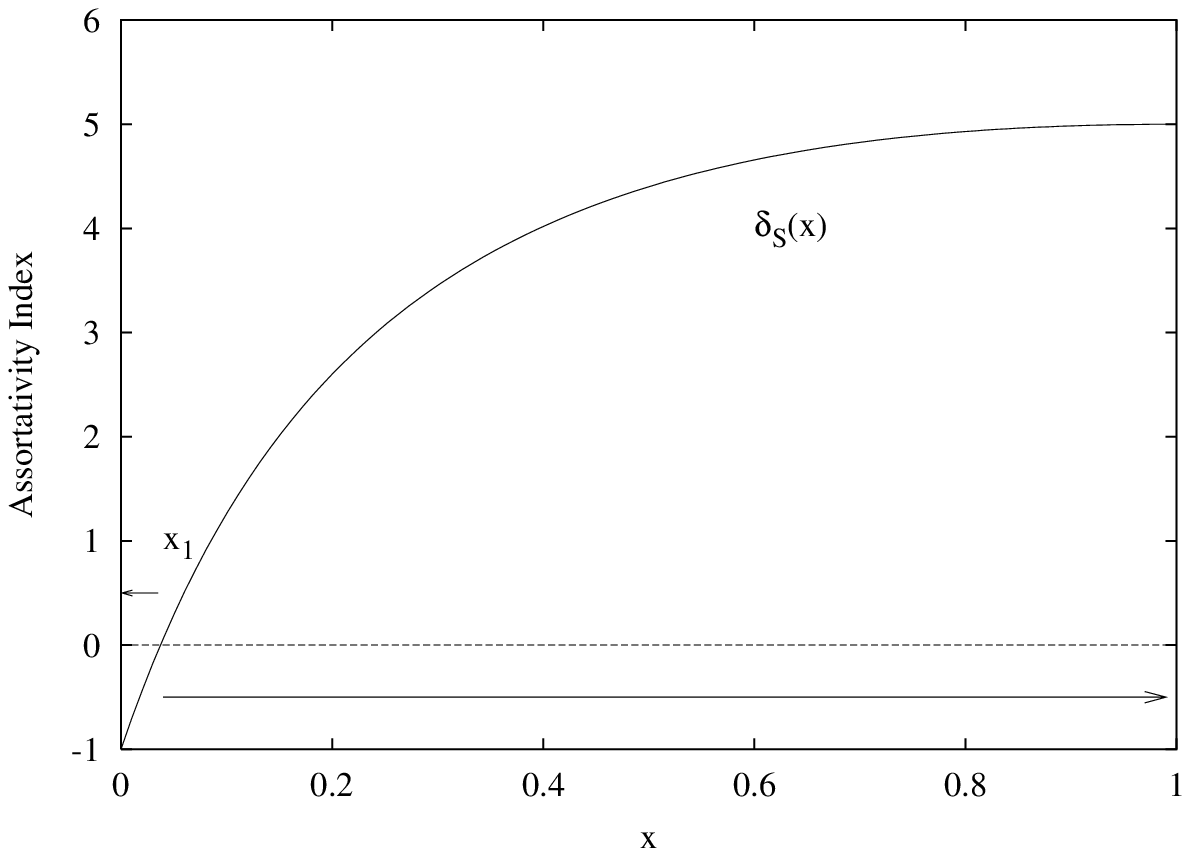}
\caption{Example of difference between payoffs $\delta(x)$ for selfknowledge model with $\alpha + \beta > 1$ . Two stable equilibria at $x =0$ and $x=1$ are present. Values used in calculation: $\alpha = 0.8$, $\beta=0.8$, $b=10$, $c=1$. Arrows indicate direction of evolution of populations. }
\end{figure}


\begin{thebibliography}{99}
\bibitem{berg1}{T. C. Bergstrom \emph{The Algebra of Assortative Encounters and the Evolution of Cooperation}}
\bibitem{maynard73}{J Maynard Smith, G. Price \emph{The Logic of Animal Conflicts}, {\bf Nature 246}:15-18, 1973}
\bibitem{maynard82} {J. Maynard Smith \emph{Evolution and the Theory of Games}, Cambridge University Press, 1982}
\bibitem{roth82} {A. Roth, M. Sotomayor \emph{Two-Sided Matching: A Study in Geme-Theoretic Modelling and Analysis}, Econometric Society Monographs, Cambridge University Press, 1982}
\end{thebibliography}
 \end{document}